\documentclass[a4paper,12pt]{article}
\usepackage{jheppub} 

\title{Electroweak phase transition
and Higgs boson couplings
in the model
based on supersymmetric strong dynamics
}

\author[a]{Shinya~Kanemura,}
\author[b]{Eibun~Senaha,}
\author[c]{Tetsuo~Shindou}
\author[d]{and Toshifumi~Yamada}

\affiliation[a]{Department of Physics, University of Toyama\\
	3190 Gofuku, Toyama 930-8555, Japan}
\affiliation[b]{School of Physics, KIAS\\ 
85 Hoegiro, Dongdaemu-gu, Seoul 130-722, Korea}
\affiliation[c]{Division of Liberal Arts, Kogakuin University\\
1-24-2 Nishi-Shinjuku, Tokyo 163-8677, Japan}
\affiliation[d]{Department of Physics, University of Tokyo\\
7-3-1 Hongo, Tokyo 113-0033, Japan}

\emailAdd{kanemu@sci.u-toyama.ac.jp}
\emailAdd{senaha@kias.re.kr}
\emailAdd{shindou@cc.kogakuin.ac.jp}
\emailAdd{toshifumi@hep-th.phys.s.u-tokyo.ac.jp}

\abstract{
We discuss a strongly-coupled extended Higgs sector with the 126 GeV Higgs boson,
 which is a low-energy effective theory of the supersymmetric SU(2)$_H$ gauge thoery 
 that causes confinement.
In this effective theory,
 we study the parameter region where 
 electroweak phase transition is of strongly first order,
 as required for successful electroweak baryogenesis.
In such a parameter region, the model has a Landau pole at the order of 10 TeV,
 which corresponds to the confinement scale of the SU(2)$_H$ gauge theory.
We find that the large coupling constant which blows up at the Landau pole
 results in large non-decoupling loop effects on low-energy observables,
 such as the Higgs-photon-photon vertex and the triple Higgs boson vertex.
As phenomenological consequences of electroweak baryogenesis in our model,
 the Higgs-to-diphoton branching ratio is about 20\% smaller 
 while the triple Higgs boson coupling is more than about 20\% larger
 than the standard model predictions.
Such deviations may be detectable in future collider experiments.
}

\preprint{\\ \rightline{UT-HET-075, KIAS-P12077, KU-PH-012, UT-12-39}}

\begin{document} 
\maketitle
\flushbottom

\section{Introduction}

\ \ \ Successful electroweak baryogenesis (EWBG)~\cite{ewbg} 
 relies on sufficient amount of CP violation and strongly first order 
 electroweak phase transition (EWPT).
In the standard model (SM), it turns out that
 the Kobayashi-Maskawa phase is far too small to generate sufficient
 baryon asymmetry~\cite{ewbg_sm_cp}, and the EWPT is a smooth crossover for a Higgs boson
 with the mass above 73 GeV~\cite{crossover}. 
Therefore, the SM must be extended.
In general, extra CP-phases naturally enter into extended Higgs models.
On the other hand, the condition on the EWPT
 is directly connected to the structure of the Higgs potential.

Many attempts to obtain feasible EWBG scenarios have been done 
 in the extended models~\cite{thdm,kos,mssm pt,Funakubo:2009eg,nmssm pt,kss,fok}.
Among them, much attention has been paid to
 the minimal supersymmetric SM (MSSM) so far.
As pointed out in refs.~\cite{lhc_light_stop}, 
 however, the light scalar top scenario that is necessary for successful EWBG in MSSM 
 is in tension with the current experimental data, 
 especially the Higgs signal strength measurements at the LHC.
Therefore, it is time to consider alternative models for successful EWBG in some detail,
 taking the recent LHC data into account.

Now that the Higgs boson mass is found to be 126 GeV,
 models that realize strongly first order phase transition
 due to a thermal cubic term generally require relatively large coupling constants in the Higgs sector.
Such large coupling constants can blow up below the Planck scale with a Landau pole.
In this case, the model
 must be replaced by a more fundamental theory at the Landau pole.
Recently,
 an ultraviolet (UV) complete framework has been proposed
 for an extended Higgs sector 
 incorporating such large coupling constants \cite{fatbutlight},
 which is based on the SUSY SU(2)$_H$ gauge theory with six doublets, $T_1, ..., T_6$ ($N_f=3$), and one singlet
 as a UV theory.
This simple gauge structure was originally applied
 to the minimal supersymmetric fat Higgs model \cite{fat higgs}.
The gauge theory is strongly-coupled at an infrared scale.
In ref.~\cite{fatbutlight}, the SU(2)$_H$ doublets $T_1, ..., T_6$ are confined to give mesonic superfields \cite{is},
 which are identified with the MSSM Higgs superfields and
 other exotic superfields in the extended Higgs sector.
The Landau pole at which the coupling constants in the extended Higgs sector blow up
 is nothing but the confinement scale of the SU(2)$_H$ gauge theory.
A striking feature of this framework is that
 the large coupling constants as well as
 the field content of the extended Higgs sector
 automatically result from the dynamics of the SUSY gauge theory.
The Landau pole is determined to be of the order of 10 TeV
 from the requirement of strongly first order EWPT \cite{kss}.

The low-energy effective theory of the model in ref.~\cite{fatbutlight}
 contains four Higgs doublet, a pair of charged singlet and a pair of neutral singlet
 chiral superfields.
In order to avoid flavor-changing neutral current,
 additional discrete $Z_2$ symmetry is imposed on the model,
 under which extra two doublets and all the singlets are odd.
When the discrete symmetry is exact,
 the lightest $Z_2$-odd particle can be another dark matter candidate 
 other than the lightest supersymmetric particle
 as long as it is electrically neutral.
Since this framework has the Landau pole around 10 TeV,
 it is unclear whether the na$\ddot{\i}$ve seesaw mechanism \cite{seesaw} for tiny neutrino masses 
 with very heavy Majorana masses can be applied or not.
A radiative generation mechanism 
 for tiny neutrino masses with $Z_2$-odd extra scalar doublets
 and $Z_2$-odd TeV scale right handed neutrinos \cite{ma, aks}
 may be compatible with our framework.
By developing the model in ref.~\cite{fatbutlight},
 we may be able to build a fundamental UV complete model
 whose low-energy effective theory can explain 
 baryogenesis, dark matter and tiny neutrino mass 
 simultaneously below its confinement scale.

In general, loop effects due to a heavy particle 
 becomes suppressed from low energy observables 
 in the large mass limit by the decoupling theorem \cite{ac}. 
However, when the mass mainly comes from the vacuum expectation value (VEV), 
 the decoupling theorem does not hold, and nonvanishing 
 loop effect can appear. 
In particular, large non-decoupling effects can appear in the triple Higgs boson coupling \cite{triple}
 and the decay branching ratio of the Higgs boson to diphoton \cite{nd-hgamgam}.
Since the enhancement of first order EWPT also comes from the non-decoupling effect,
 the strength of first order EWPT and those low-energy observables
 are correlated.
\footnote{
 The correlation between the strength of first order EWPT and the triple Higgs boson coupling 
  has been discussed in different models in ref.~\cite{perelstein},
  and that between the strength of first order EWPT and the Higgs-to-diphoton branching ratio
  has been discussed in a different model in ref.~\cite{davoudiasl}.
}
In fact, significant deviations of those observables from the SM predictions
 are found in the two Higgs doublet model when EWPT is of strongly first order \cite{kos}.
Also in SUSY extended Higgs models such as the ``four Higgs doublets + two charged singlets model",
 it is possible to realize strongly first order EWPT by large non-decoupling loop effects \cite{kss},
 and at the same time large non-decoupling effects contribute to low-energy observables \cite{ksy'}.
Since these models contain a Landau pole,
 it is unclear how these models are related to physics at UV scales.
In addition, the correlation between
 the strength of EWPT and low-energy observables has not been properly studied 
 based on UV complete models.

In this paper,
 we discuss phenomenology of the extended Higgs model that emerges
 as a low-energy effective theory of the SUSY SU(2)$_H$ gauge theory with six doublets and one singlet,
 proposed in ref.~\cite{fatbutlight}.
First we evaluate the strength of EWPT 
 and seek for parameter regions where strongly first order EWPT occurs,
 as is necessary for successful EWBG.
We then calculate, in such parameter regions,
 the Higgs-to-diphoton branching ratio and the triple Higgs boson coupling,
 the former of which is measurable by 5\% accuracy \cite{peskin}
 and the latter of which by 20\% accuracy \cite{fujii}
 in future collider experiments.
The relationship among the strength of EWPT and these two quantities
 is investigated.

This paper is organized as follows.
In the next section, we give a detailed description of
 the extended Higgs model with large coupling constants
 that originates from the SUSY SU(2)$_H$ gauge theory.
In Section 3, we evaluate the strength of EWPT.
In Section 4, we summarize the calculation of the decay branching ratio
 of the Higgs boson into diphoton and the triple Higgs boson coupling.
In Section 5, we take a benchmark mass spectrum,
 and discuss the correlation among the strength of EWPT,
 the Higgs-to-diphoton branching ratio and the triple Higgs boson coupling
 in our extended Higgs model.
The final section is devoted to conclusions.
\\
\\

\section{Model}

\subsection{Lagrangian}

\ \ \ We consider a supersymmetric (SUSY) extended Higgs sector 
 that emerges as a low-energy effective theory of
 the SUSY SU(2)$_H$ gauge theory with three pairs of doublets and one singlet, 
 which has been proposed in ref.~\cite{fatbutlight}.
In this model, the mesonic superfields of the SU(2)$_H$ gauge theory
 are identified with the Higgs doublets of the MSSM as well as
 the extra chiral superfields in the extended Higgs sector.
We stress that
 the field content and the superpotential of the model are uniquely determined by
 the dynamics of the gauge theory.

The model contains
 two SU(2)$_L$ doublet, two charged singlet and four neutral singlet chiral superfields,
 in addition to the two Higgs doublets of the MSSM.
The model has a $Z_2$ parity, under which the MSSM fields and two neutral singlets are even
 and the others are odd.
This parity forbids the Yukawa couplings between the extra SU(2)$_L$ doublets and matter superfields
 so that dangerous flavor changing neutral current processes are suppressed.
The field content of the Higgs sector is summarized in Table 1.
\begin{table}
\begin{center}
\begin{tabular}{|c|c|c|c|} \hline
Field                 & $SU(2)_{L}$ & $U(1)_{Y}$ & $Z_{2}$ \\ \hline
$H_{u}$               & 2           & +1/2       & $+$      \\ \hline
$H_{d}$               & 2           & $-$1/2       & $+$       \\ \hline
$\Phi_{u}$            & 2           & +1/2       & $-$       \\ \hline
$\Phi_{d}$            & 2           & $-$1/2       & $-$       \\ \hline
$\Omega^+$               & 1           & +1       & $-$      \\ \hline
$\Omega^-$               & 1           & $-$1       & $-$       \\ \hline
$\zeta$, $\eta$          & 1           & 0       & $-$       \\ \hline
$n_{\Phi}$, $n_{\Omega}$ & 1           & 0       & $+$       \\ \hline
\end{tabular}
\end{center}
\caption{
Properties of the fields in the Higgs sector under the SM gauge groups and the $Z_2$ parity.
}
\end{table}
The superpotential of the Higgs sector is given by
\begin{eqnarray}
W_{Higgs} &=& -\mu (H_uH_d-n_{\Phi}n_{\Omega})-\mu_{\Phi}\Phi_u\Phi_d-\mu_{\Omega}(\Omega^+\Omega^- - \zeta\eta)
	\nonumber\\
	&+&\hat{\lambda} \ \left\{ \
	H_d\Phi_u\zeta + H_u\Phi_d\eta - H_u\Phi_u\Omega^- - H_d\Phi_d\Omega^+
	+n_{\Phi}\Phi_u\Phi_d +n_{\Omega}(\Omega^+\Omega^- -\zeta\eta) \ \right\} \label{superpot} \ .
\nonumber \\
\end{eqnarray}
$\hat{\lambda}$ denotes a running coupling constant for the fields in the extended Higgs sector.
As for the superfields $n_{\Phi}$ and $n_{\Omega}$,
 their scalar components couple to the MSSM Higgs scalars at tree level.
However these couplings do not contribute to the one-loop effective potential for the MSSM Higgs scalars.
We therefore ignore $n_{\Phi}$ and $n_{\Omega}$ in the following discussion.
The soft SUSY breaking terms are introduced as follows:
\begin{eqnarray}
{\cal L}_{soft} &=& -m_{H_u}^2 H_u^{\dagger} H_u \ - \ m_{H_d}^2 H_d^{\dagger} H_d
        \ - \ m_{\Phi_u}^2 \Phi_u^{\dagger} \Phi_u \ - \ m_{\Phi_d}^2 \Phi_d^{\dagger} \Phi_d \nonumber
\\
        &-& m_{\Omega^+}^2 \Omega^{+ \, \dagger} \Omega^+ \ - \ m_{\Omega^-}^2 \Omega^{- \, \dagger} \Omega^-
        \ - \ m_{\zeta}^2 \zeta^{\dagger} \zeta \ - \ m_{\eta}^2 \eta^{\dagger} \eta \nonumber
\\
&-& B\mu H_u H_d \ - \ B\mu_{\Phi} \Phi_u \Phi_d \ - \ B\mu_{\Omega} ( \Omega^+ \Omega^- - \zeta \eta) \nonumber
\\
&-& A_{\zeta} H_d\Phi_u\zeta \ - \ A_{\eta} H_u\Phi_d\eta 
\ - \ A_{\Omega^-} H_u\Phi_u\Omega^- \ - \ A_{\Omega^+} H_d\Phi_d\Omega^+ \ .
\end{eqnarray}
\\

\subsection{Mass Matrices}

\ \ \ We denote the VEV of $H_u^0$ by $v_u/\sqrt{2}$ and that of $H_d^0$ by $v_d/\sqrt{2}$.
We hereafter denote the value of the running coupling constant, $\hat{\lambda}$, at the electroweak scale
 by $\lambda \equiv \hat{\lambda}(\mu_{EW}^{})$.

The scalar and fermionic components of $H_u$ and $H_d$ have 
 the same mass spectrum as MSSM at tree level.
The fermionic components of the $Z_2$-odd superfields
 have the following mass matrices at tree level:
\begin{eqnarray}
{\cal L}_{{\rm odd \ charginos}} &=& 
-\left(
\begin{array}{cc}
\tilde{\Phi}_u^+, & \tilde{\Omega}^+
\end{array}
\right)
\left(
\begin{array}{cc}
-\mu_{\Phi} & \lambda v_u / \sqrt{2} \\
-\lambda v_d / \sqrt{2} & -\mu_{\Omega}
\end{array}
\right)
\left(
\begin{array}{c}
\tilde{\Phi}_d^- \\
\tilde{\Omega}^-
\end{array}
\right) \\
{\cal L}_{{\rm odd \ neutralinos}} &=& -\frac{1}{2} \ 
\left(
\begin{array}{cccc}
\tilde{\Phi}_u^0, & \tilde{\Phi}_d^0, & \tilde{\zeta}, & \tilde{\eta}
\end{array}
\right)
\left(
\begin{array}{cccc}
0 & \mu_{\Phi} & \lambda v_d / \sqrt{2} & 0 \\
\mu_{\Phi} & 0 & 0 & -\lambda v_u / \sqrt{2} \\
\lambda v_d / \sqrt{2} & 0 & 0 & \mu_{\Omega} \\
0 & -\lambda v_u / \sqrt{2} & \mu_{\Omega} & 0
\end{array}
\right)
\left(
\begin{array}{c}
\tilde{\Phi}_u^0 \\
\tilde{\Phi}_d^0 \\
\tilde{\zeta} \\
\tilde{\eta}
\end{array}
\right) \nonumber \\
\end{eqnarray}
The scalar components of the $Z_2$-odd superfields
 have the following mass matrices at tree level:
\begin{eqnarray}
& & {\cal L}_{{\rm odd \ charged \ scalars}} \ = \ 
-\left(
\begin{array}{cccc}
(\Phi_u^+)^*, & (\Omega^+)^*, & \Phi_d^-, & \Omega^-
\end{array}
\right) \times \nonumber
\\ 
& &
\left(
\begin{array}{cccc}
\bar{m}_{\Phi_u}^2 + \lambda^2 \frac{v_u^2}{2} + D_{\Phi \pm} & \lambda \mu_{\Phi}^* \frac{v_d}{\sqrt{2}} - \lambda \frac{v_u}{\sqrt{2}} \mu_{\Omega} & B \mu_{\Phi}^* & \lambda \mu \frac{v_d}{\sqrt{2}} - A_{\Omega^-}^* \frac{v_u}{\sqrt{2}} \\
\lambda \mu_{\Phi} \frac{v_d}{\sqrt{2}} - \lambda \mu_{\Omega}^* \frac{v_u}{\sqrt{2}} & \bar{m}_{\Omega^+}^2 + \lambda^2 \frac{v_d^2}{2} + D_{\Omega \pm} & -\lambda \mu \frac{v_u}{\sqrt{2}} + A_{\Omega^+}^* \frac{v_d}{\sqrt{2}} & B \mu_{\Omega}^* \\
B \mu_{\Phi} & -\lambda \mu^* \frac{v_u}{\sqrt{2}} + A_{\Omega^+} \frac{v_d}{\sqrt{2}} & \bar{m}_{\Phi_d}^2 + \lambda^2 \frac{v_d^2}{2} - D_{\Phi \pm} & \lambda \mu_{\Omega}^* \frac{v_d}{\sqrt{2}} - \lambda \mu_{\Phi} \frac{v_u}{\sqrt{2}} \\
\lambda \mu^* \frac{v_d}{\sqrt{2}} - A_{\Omega^-}^* \frac{v_u}{\sqrt{2}} & B \mu_{\Omega} & \lambda \mu_{\Omega} \frac{v_d}{\sqrt{2}} - \lambda \mu_{\Phi}^* \frac{v_u}{\sqrt{2}} & \bar{m}_{\Omega^-}^2 + \lambda^2 \frac{v_u^2}{2} - D_{\Omega \pm}
\end{array}
\right) \nonumber
\\
&\times&
\left(
\begin{array}{c}
\Phi_u^+ \\
\Omega^+ \\
(\Phi_d^-)^* \\
(\Omega^-)^*
\end{array}
\right) \label{chsc}
\\
& & {\cal L}_{{\rm odd \ neutral \ scalars}} \ = \ 
-\left(
\begin{array}{cccc}
(\Phi_u^0)^*, & \zeta, & \Phi_d^0, & (\eta)^*
\end{array}
\right) \times \nonumber
\\ 
& &
\left(
\begin{array}{cccc}
\bar{m}_{\Phi_u}^2 + \lambda^2 \frac{v_d^2}{2} + D_{\Phi 0} & \lambda \mu \frac{v_u}{\sqrt{2}} + A_{\zeta}^* \frac{v_d}{\sqrt{2}} & B \mu_{\Phi}^* & \lambda \mu_{\Omega} \frac{v_d}{\sqrt{2}} - \lambda \mu_{\Phi}^* \frac{v_u}{\sqrt{2}} \\
\lambda \mu^* \frac{v_u}{\sqrt{2}} + A_{\zeta} \frac{v_d}{\sqrt{2}} & \bar{m}_{\zeta}^2 + \lambda^2 \frac{v_d^2}{2} & \lambda \mu_{\Phi}^* \frac{v_d}{\sqrt{2}} - \lambda \mu_{\Omega} \frac{v_u}{\sqrt{2}} & B \mu_{\Omega} \\
B \mu_{\Phi} & \lambda \mu_{\Phi} \frac{v_d}{\sqrt{2}} - \lambda \mu_{\Omega}^* \frac{v_u}{\sqrt{2}} & \bar{m}_{\Phi_d}^2 + \lambda^2 \frac{v_u^2}{2} - D_{\Phi 0} & -\lambda \mu^* \frac{v_d}{\sqrt{2}} - A_{\eta} \frac{v_u}{\sqrt{2}} \\
\lambda \mu_{\Omega}^* \frac{v_d}{\sqrt{2}} - \lambda \mu_{\Phi} \frac{v_u}{\sqrt{2}} & B \mu_{\Omega}^* & -\lambda \mu \frac{v_d}{\sqrt{2}} - A_{\eta}^* \frac{v_u}{\sqrt{2}} & \bar{m}_{\eta}^2 + \lambda^2 \frac{v_u^2}{2}
\end{array}
\right) \nonumber
\\
&\times&
\left(
\begin{array}{c}
\Phi_u^0 \\
(\zeta)^* \\
(\Phi_d^0)^* \\
\eta
\end{array}
\right) \label{nesc}
\end{eqnarray}
In eqs.~(\ref{chsc}) and (\ref{nesc}),
 $\bar{m}_{\Phi_{u/d}}^2$, $\bar{m}_{\Omega^{\pm}}^2$, $\bar{m}_{\zeta}^2$ and $\bar{m}_{\eta}^2$
 are defined as
\begin{eqnarray}
\bar{m}_{\Phi_{u/d}}^2 &\equiv& \vert \mu_{\Phi} \vert^2 \ + \ m_{\Phi_{u/d}}^2 \ , \nonumber \\
\bar{m}_{\Omega^{\pm}}^2 &\equiv& \vert \mu_{\Omega} \vert^2 \ + \ m_{\Omega^{\pm}}^2 \ , \nonumber \\
\bar{m}_{\zeta}^2 &\equiv& \vert \mu_{\Omega} \vert^2 \ + \ m_{\zeta}^2 \ , \nonumber \\
\bar{m}_{\eta}^2 &\equiv& \vert \mu_{\Omega} \vert^2 \ + \ m_{\eta}^2 \ ,
\end{eqnarray}
 and $D_{\Phi \pm}$, $D_{\Omega \pm}$ and $D_{\Phi 0}$ represent D-term contributions 
 that are given by
\begin{eqnarray}
D_{\Phi \pm} &=& \frac{g_Y^2}{8} \ ( \ v_u^2 - v_d^2 \ ) \ - \ \frac{g^2}{8} \ ( \ v_u^2 - v_d^2 \ ) \ , \nonumber
\\
D_{\Omega \pm} &=& \frac{g_Y^2}{2} \ ( \ v_u^2 - v_d^2 \ ) \ , \nonumber
\\
D_{\Phi 0} &=& \frac{g_Y^2}{8} \ ( \ v_u^2 - v_d^2 \ ) \ + \ \frac{g^2}{8} \ ( \ v_u^2 - v_d^2 \ ) \ ,
\end{eqnarray}
 where $g_Y$ and $g$ respectively denote the gauge couplings of U(1)$_Y$ 
 (normalized so that $H_u$ has charge $Y=+1/2$) and SU(2)$_L$.

We define $m_{\Phi_1^{\prime \, 0}}^2$ and $m_{\Phi_1^{\prime \, \pm}}^2$ as 
 the smallest eigenvalues of the $Z_2$-odd neutral scalar mass matrix, eq.~(\ref{chsc}),
 and the $Z_2$-odd neutral scalar mass matrix, eq.~(\ref{nesc}), respectively.
\\

\begin{figure}[htbp]
 \begin{minipage}{0.5\hsize}
  \begin{center}
   \includegraphics[width=80mm]{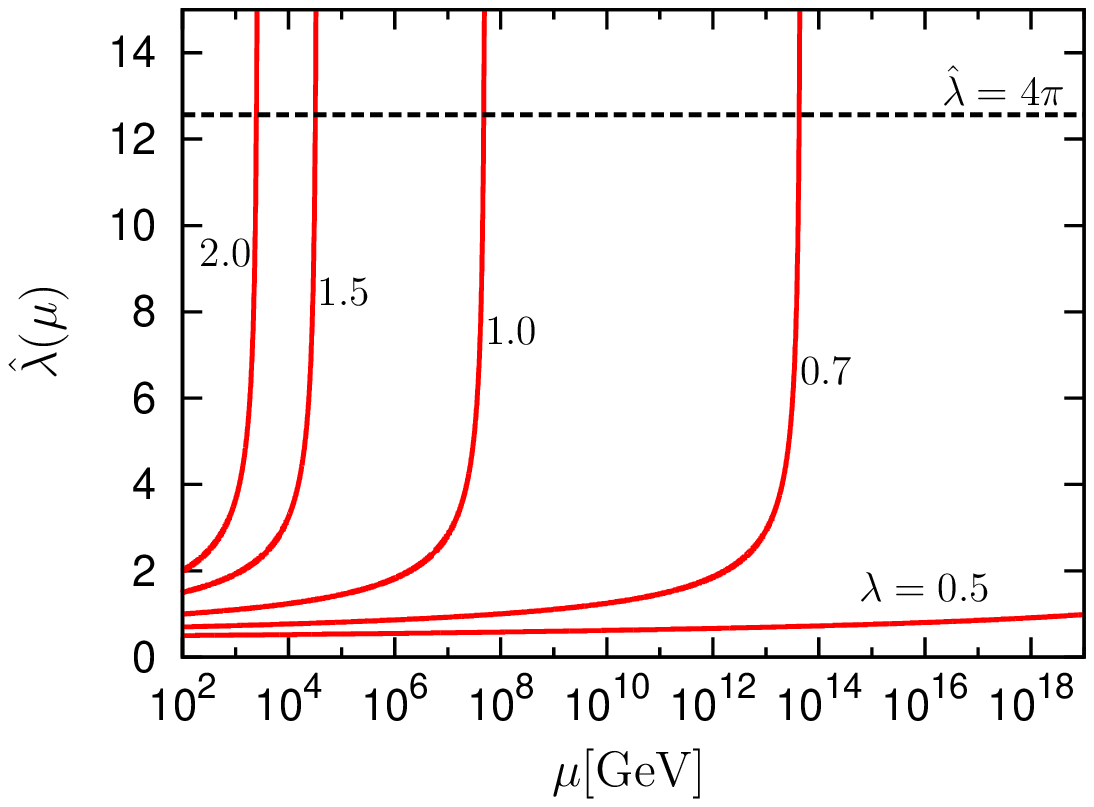}
  \end{center}
 \end{minipage}
 \begin{minipage}{0.5\hsize}
  \begin{center}
   \includegraphics[width=80mm]{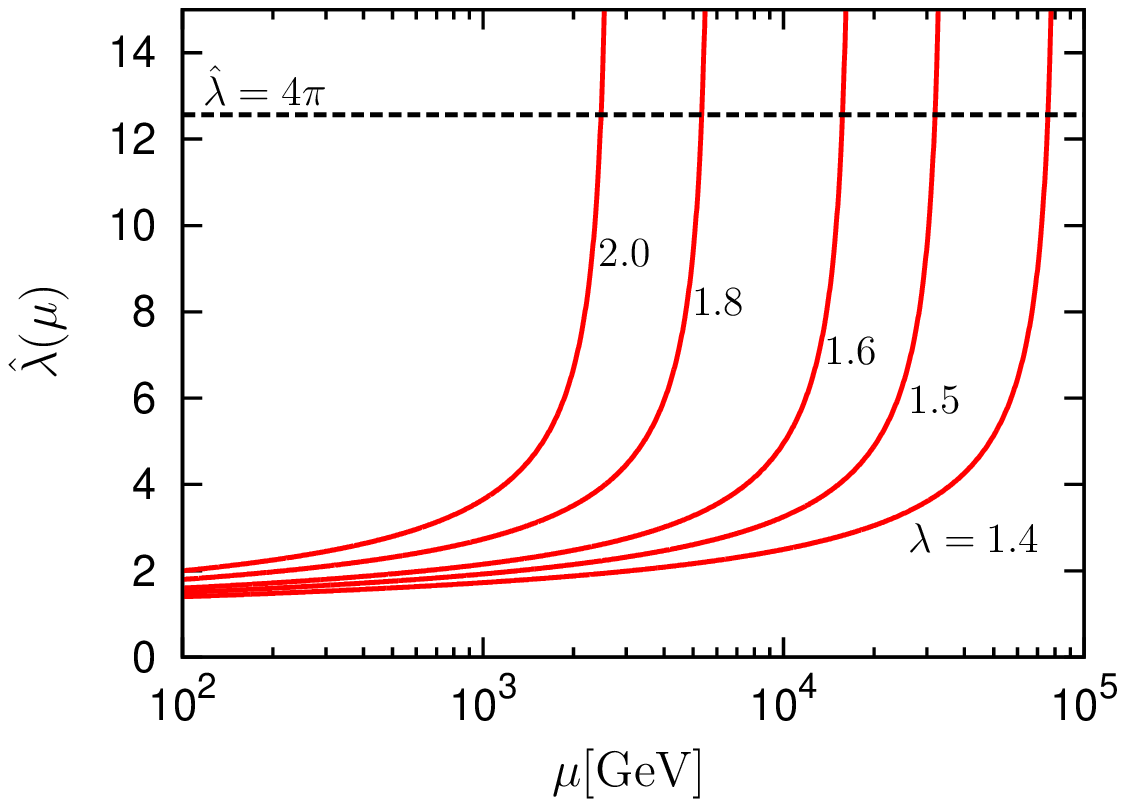}
  \end{center}
 \end{minipage}
  \caption{The renormalization group running of the coupling constant $\hat{\lambda}$
  for each value of $\hat{\lambda}$ at the electroweak scale, $\lambda \equiv \hat{\lambda}(\mu_{EW}^{})$.
  }
\label{figlam}
\end{figure}

\begin{figure}[htbp]
  \begin{center}
   \includegraphics[width=100mm]{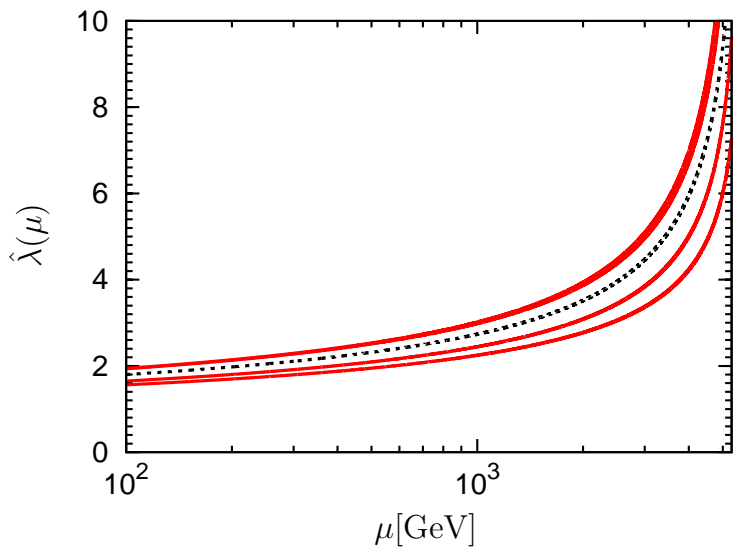}
  \end{center}
  \caption{
        The renormalization group running of the coupling constants.
	The cut-off scale $\Lambda_H$ is taken as $\Lambda_H=5.3\text{GeV}$ which is 
	relevant to $\lambda(\mu_{\text{EW}})=1.8$ in the universal coupling constant case.
	The dotted (black) line corresponds to the case with the universal coupling constant 
	for the terms in the second line of eq.~(\ref{superpot}).
	The solide (red) lines are for the case (non-universal coupling constants case)
	when the first four terms in the second line of eq.~(\ref{superpot}) have different coupling constants 
	at the cut-off scale $\Lambda_H$ 
	as $\hat{\lambda}(\Lambda_H)= 1.4\times 4\pi, 1.2\times 4\pi, 0.8\times 4\pi, 0.6\times 4\pi$.
  }
  \label{non-universal}
\end{figure}

\subsection{Coupling Constants}

\ \ \ The superpotential, eq.~(\ref{superpot}), emerges as an effective theory
 below the confinement scale of the SU(2)$_H$ gauge theory.
The running coupling constant, $\hat{\lambda}$, in the superpotential is estimated in the following way.
At the confinement scale of the SU(2)$_H$ gauge theory, $\Lambda_H$,
 SUSY Na$\ddot{\i}$ve Dimensional Analysis \cite{nda} suggests
\begin{eqnarray}
\hat{\lambda}(\Lambda_H) &\simeq& 4 \pi \label{nda} \ .
\end{eqnarray}
Below $\Lambda_H$, $\hat{\lambda}$ obeys the following renormalization group equation:
\begin{eqnarray}
\mu \frac{{\rm d} \hat{\lambda}}{{\rm d} \mu} &\simeq& \frac{6}{16 \pi^2} \ \hat{\lambda}^3 \label{rge} \ , 
\end{eqnarray}
 where we neglect the electroweak gauge couplings.
Conversely, once we know the value of the coupling constant $\hat{\lambda}$ at the electroweak scale,
 we can determine the confinement scale $\Lambda_H$ by using eqs.~(\ref{nda}) and (\ref{rge}).
Figure~\ref{figlam} describes the renormalization group running of $\hat{\lambda}$
 for each value of $\hat{\lambda}$ at the electroweak scale, $\lambda \equiv \hat{\lambda}(\mu_{EW}^{})$.
For example, if $\lambda=1.6$ ($\lambda=1.8)$,
 the confinement scale exists around 15 TeV (5 TeV).
We note that the relation between the value of $\lambda$ and the confinement scale $\Lambda_H$
 is rather robust
 even though the estimate eq.~(\ref{nda}) is subject to $O(1)$ ambiguity,
 because the running coupling constant $\hat{\lambda}(\mu)$ shows a steep rise near the confinement scale.
We also note that,
 if the coupling constants at the scale $\Lambda_H$ are not universal and differ by the factor of 2,
 those at the electroweak scale differ at most by the factor of 0.1.
As an example, we show in Figure~(\ref{non-universal}) 
 the renormalization group running of the coupling constants 
 for the case when the confinement scale is 5.3 TeV and 
 the four coupling constants for the terms 
 $H_d \Phi_u \zeta$, $H_u \Phi_d \eta$, $H_u \Phi_u \Omega^-$, $H_d \Phi_d \Omega^+$
 in eq.~(\ref{superpot}) take the values of
 1.4$\times 4 \pi$, 1.2$\times 4 \pi$,  0.8$\times 4 \pi$, 0.6$\times 4 \pi$ at that scale.


\section{Electroweak Phase Transition}

\ \ \ In the EWBG scenario, the BAU is created in the symmetric phase
where the $(B+L)$-changing processes are active.
In order to leave the generated BAU as it is, 
the $(B+L)$-changing rate in the broken phase must be sufficiently suppressed.
The conventional criterion is
\begin{eqnarray}
\frac{v_c}{T_c}&\gtrsim& C\ ,\label{sph_dec}
\end{eqnarray}
where $T_c$ denotes a critical temperature 
at which the effective potential has two degenerate minima, 
$v_c=\sqrt{v_d^2(T_c)+v_u^2(T_c)}$, and $C$ is a parameter that depends
on the sphaleron energy and so on.
In this paper, we simply take $C\simeq 1$ rather than evaluating the precise value.
In the MSSM, $C\simeq 1.4$~\cite{Funakubo:2009eg}.
As demonstrated in ref.~\cite{kss}, $v_c$ and $T_c$ are determined using the 
one-loop effective potentials with ring resummations.

\begin{figure}[t]
\center
\includegraphics[width=7.2cm]{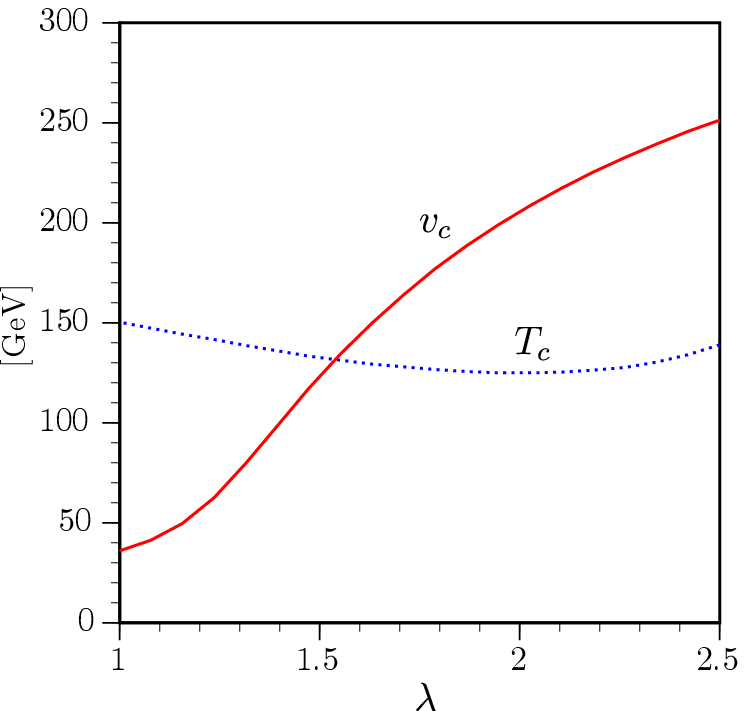} \hspace{0.1cm}
\includegraphics[width=7.2cm]{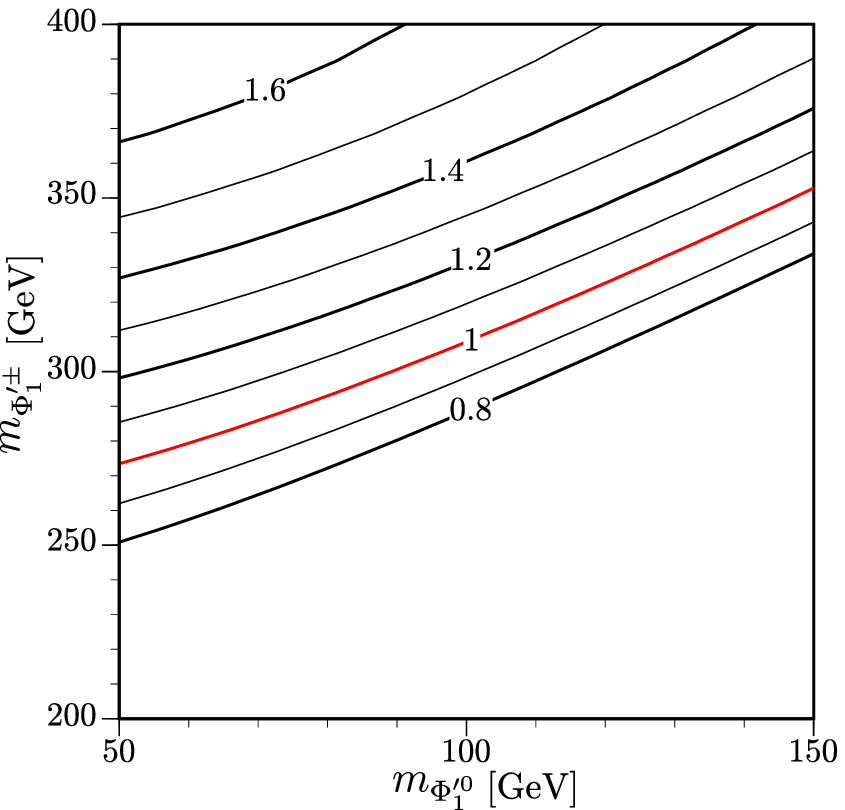}
\caption{(Left panel) $v_c$ and $T_c$ as a function of $\lambda$. 
(Right panel) The contour of $v_c/T_c$ in $m_{\Phi'^\pm_1}$-$m_{\Phi'^0_1}$ plane.
The input parameters are given in the text.}
\label{fig:vc-Tc-lam}
\end{figure}

In the left panel of Figure~\ref{fig:vc-Tc-lam}, $v_c$ and $T_c$ are shown as a function of $\lambda$.
We take $\tan\beta=15$, $m_{H^\pm}=350$ GeV, 
$\bar{m}_{\Omega^+}^2=\bar{m}_{\Phi_d}^2=\bar{m}_\zeta^2=(1500~{\rm GeV})^2$, 
$\bar{m}_\eta^2=(2000~{\rm GeV})^2$, 
$\bar{m}_{\Omega^-}^2=\bar{m}_{\Phi_u}^2=(50~{\rm GeV})^2$, 
$B_\Omega=B_\Phi=0$ and $\mu_\Phi=-\mu_\Omega=550$ GeV.
We tune the A-term of the top squarks to realize 
 126 GeV mass for the SM-like Higgs boson.
Similar to the results in ref.~\cite{kss}, $\lambda\gtrsim 1.6$ leads to $v_c/T_c\gtrsim1$, 
which is due to the nondecoupling effects
arising from the $Z_2$-odd charged Higgs boson loops. 
Note that the small $\bar{m}_{\Omega^-}^2$ and $\bar{m}_{\Phi_u}^2$ are necessary to realize
such loop effects.

In the right panel of Figure~\ref{fig:vc-Tc-lam}, the contours of $v_c/T_c$ are plotted.
Here, $m_{\Phi'^\pm_1}$ and $m_{\Phi'^0_1}$ are treated as the input parameters, and 
$\bar{m}_{\Omega^-}^2$ and $\lambda$ are derived quantities.
The value of $m_{\Phi'^0_1}$ more or less fixes $\bar{m}_{\Omega^-}^2$,
and so the size of $m_{\Phi'^\pm_1}$ is mostly controlled by $\lambda v_u$.
Therefore, the nondecoupling loop effects would be strengthen as $m_{\Phi'^0_1}$ decreases
and $m_{\Phi'^\pm_1}$ increases. The plot clearly shows that $v_c/T_c$ gets enhanced 
in such a nondecoupling region. This example indicates that $m_{\Phi'^\pm_1}\gtrsim270~{\rm GeV}$ 
with a relatively light $\Phi'^0_1$ are required to be consistent with successful electroweak baryogenesis.
 \\
\\

\section{Low-energy Observables}

\ \ \ In this section, we summarize the methods to evaluate the two
 experimentally observable quantities, 
 the branching ratio of the SM-like Higgs boson into diphoton
 and the triple Higgs boson coupling
 (at zero temperature),
 and their deviations from the SM predictions.
\\

\subsection{Decay Branching Ratio of the Higgs Boson into Diphoton}

\ \ \ In our model, $Z_2$-odd charged bosons and fermions
 alter the branching ratio of the SM-like Higgs boson into diphoton, $Br(h \rightarrow \gamma \gamma)$.
Such effects can be significantly large 
 because $h \rightarrow \gamma \gamma$ process arises only at loop levels,
 and the loop diagrams involving $Z_2$-odd charged bosons and fermions
 are enhanced by their large coupling constant $\lambda$ with MSSM Higgs superfields.
The amplitude for the one-loop diagram involving the $Z_2$-odd charged bosons
 is given by \cite{hgamgam}
\begin{eqnarray}
{\cal A}_S &=& C_0 \ \sum_{i=1,..,4} \ s_i \ \lambda^2 \ \frac{v^2}{\sqrt{2} m_{\phi^{\pm}_i}^2} \ 
\frac{1}{x_{\phi^{\pm}_i}^2} \{ -x_{\phi^{\pm}_i} + f(x_{\phi^{\pm}_i}) \} \ ,
\end{eqnarray}
 and that for the one-loop diagram involving the $Z_2$-odd charged fermions
 is given by
\begin{eqnarray}
{\cal A}_F &=& C_0 \ \sum_{j=1,2} \ f_j \ \lambda^2 \ \frac{v}{m_{\chi^{\pm}_j}} \ 
\frac{2}{x_{\chi^{\pm}_j}^2} \{ x_{\chi^{\pm}_j} + (x_{\chi^{\pm}_j}-1) f(x_{\chi^{\pm}_j}) \} \ ,
\end{eqnarray}
 where $C_0$ is a common constant,
 $x_{\phi^{\pm}_i}$ and $x_{\chi^{\pm}_j}$ are respectively defined as
\begin{eqnarray}
 x_{\phi^{\pm}_i} &\equiv& m_h^2 / 4 m_{\phi^{\pm}_i}^2 \ , \ \ \ 
 x_{\chi^{\pm}_j} \ \equiv \ m_h^2 / 4 m_{\chi^{\pm}_j}^2 \ ,
\end{eqnarray}
and the function $f(x)$ is defined as
\begin{eqnarray}
f(x) &\equiv& \arcsin^2 (\sqrt{x})
\end{eqnarray}
 for $x \leq 1$.
Here $\phi^{\pm}_i$ $(i=1,2,3,4)$ denote the four mass eigenstates of the $Z_2$-odd charged scalars
 and $\chi^{\pm}_j$ $(j=1,2)$ denote the two mass eigenstates of the $Z_2$-odd charged fermions.
The coefficients $s_i$ denote the couplings between the charged scalars $\phi_i^{\pm}$ and the Higgs boson $h$
 normalized by $\lambda^2 v$,
 and $f_j$ denote those between the charged fermions $\chi_j^{\pm}$ and $h$
 normalized by $\lambda$.
We note that $s_i$ and $f_j$ depend on the mixings in the Higgs sector in a complicated way,
 but at least two of $s_i$'s and one of $f_j$'s are of order 1.

The contributions from one-loop diagrams involving the MSSM charged Higgs boson and charginos
 are negligible compared to those from $Z_2$-odd fields
 because they are suppressed by 
 the ratios of the electroweak gauge couplings over $\lambda$.
The ratio of $Br(h \rightarrow \gamma \gamma)$ over its SM value, $\mu_{\gamma \gamma}$, is written as
\begin{eqnarray}
\mu_{\gamma \gamma} &\equiv& \frac{Br(h \rightarrow \gamma \gamma)}{Br(h \rightarrow \gamma \gamma) \vert_{SM}}
 \ = \ \frac{ \vert {\cal A}_t + {\cal A}_W + {\cal A}_S + {\cal A}_F \vert^2 }{ \vert {\cal A}_t + {\cal A}_W \vert^2 } \ .
\end{eqnarray}
\\

\subsection{Triple Higgs Boson Coupling}

\ \ \ Radiative corrections due to $Z_2$-odd bosons and fermions
 affect the (zero temperature) triple Higgs boson coupling
 and causes its deviation from the SM value.
The deviation can be drastically large because of the large coupling $\lambda$
 between $Z_2$-odd superfields and MSSM Higgs superfields \cite{ksy'}.

We evaluate the triple Higgs boson coupling
 by using the one-loop effective potential \cite{cw}.
In our model, assuming that the $Z_2$-parity is not spontaneously broken,
 the one-loop effective potential is given by
\begin{eqnarray}
V_{{\rm 1-loop}} [h_u, h_d, a_u, a_d] &=& \frac{1}{64 \pi^2} \left\{ \ 
g_s \sum_{i=1}^4 m_{\phi^0_i}^4 (h_u,h_d,a_u,a_d) \left[ \log \frac{ m_{\phi^0_i}^2(h_u,h_d,a_u,a_d) }{ Q^2 } - \frac{3}{2} \right] \right.\nonumber
\\
&+& g_s \sum_{i=1}^4 \ m_{\phi^{\pm}_i}^4 (h_u,h_d,a_u,a_d) \ \left[ \ \log \frac{ m_{\phi^{\pm}_i}^2(h_u,h_d,a_u,a_d) }{ Q^2 } \ - \ \frac{3}{2} \ \right] \nonumber
\\
&-& g_M \sum_{i=1}^4 \ m_{\chi^0_i}^4 (h_u,h_d,a_u,a_d) \ \left[ \ \log \frac{ m_{\chi^0_i}^2(h_u,h_d,a_u,a_d) }{ Q^2 } \ - \ \frac{3}{2} \ \right] \nonumber
\\
&-& \left. g_D \sum_{i=1}^2 \ m_{\chi^{\pm}_i}^4 (h_u,h_d,a_u,a_d) \ \left[ \ \log \frac{ m_{\chi^{\pm}_i}^2(h_u,h_d,a_u,a_d) }{ Q^2 } \ - \ \frac{3}{2} \ \right] \ \right\} \ ,
\nonumber \\
\end{eqnarray}
 where $Q$ corresponds to a renormalization scale,
 $m_{\phi^0_i}$, $m_{\phi^{\pm}_i}$, $m_{\chi^0_i}$ and $m_{\chi^{\pm}_i}$
 respectively denote the mass eigenvalues of $Z_2$-odd neutral scalars, charged scalars,
 neutral Majorana fermions and charged Dirac fermions
 which depend on the values of the neutral components of the MSSM Higgs bosons,
\begin{eqnarray}
\langle H_u^0 \rangle &=& \frac{h_u}{\sqrt{2}} \ + \ i \frac{a_u}{\sqrt{2}} \ , \ \ \ \langle H_d^0 \rangle \ = \ \frac{h_d}{\sqrt{2}} \ + \ i \frac{a_d}{\sqrt{2}} \ ,
\end{eqnarray}
 and $g_s$, $g_M$ and $g_D$ respectively count the physical degrees of freedom
 of a complex scalar, a Majorana fermion and a Dirac fermion,
 and are given as $g_s=2$, $g_M=2$ and $g_D=4$.

The Higgs potential at one-loop level is written as
\begin{eqnarray}
V[h_u,h_d,a_u,a_d] &=& V_{{\rm tree}} \ + \ V_{{\rm 1-loop}} \ ,
\end{eqnarray}
 where $V_{{\rm tree}}$ denotes the tree level potential.
The mass eigenstates $h$, $H$ and $A$ as well as the Nambu-Goldstone mode $G$ are related to
 $h_u$, $h_d$, $a_u$ and $a_d$ by
\begin{eqnarray}
\left(
\begin{array}{c}
h_u \\
h_d
\end{array}
\right)
&=&
\left(
\begin{array}{cc}
\cos \alpha & \sin \alpha \\
-\sin \alpha & \cos \alpha
\end{array}
\right)
\left(
\begin{array}{c}
h \\
H
\end{array}
\right) \ ,
\\
\left(
\begin{array}{c}
a_u \\
a_d
\end{array}
\right)
&=&
\left(
\begin{array}{cc}
\sin \beta & \cos \beta \\
-\cos \beta & \sin \beta
\end{array}
\right)
\left(
\begin{array}{c}
G \\
A
\end{array}
\right) \ .
\end{eqnarray}
We here choose as a set of input parameters
 $v (\equiv \sqrt{v_u^2+v_d^2} \simeq 246$ GeV), $\tan \beta (\equiv v_u/v_d)$, 
 $m_h, \ m_H, \ m_A$, where $m_{\varphi}$ denotes the mass of $\varphi$,
 and the mixing angle of the CP-even Higgs bosons $\alpha$.
We impose the following renormalization conditions.
At $h_u = v \sin \beta (=v_u), \ h_d = v \cos \beta (=v_d)$ and $a_u=a_d=0$,
\begin{eqnarray}
\frac{\partial V}{\partial h} &=& \frac{\partial V}{\partial H} \ = \ 0 \ ,
\\
\frac{\partial^2 V}{\partial h^2} &=& m_h^2 \ , \ \ \ \ \
\frac{\partial^2 V}{\partial H^2} \ = \ m_H^2 \ , \ \ \ \ \ 
\frac{\partial^2 V}{\partial H \partial h} \ = \ 0 \ , \ \ \ \ \
\frac{\partial^2 V}{\partial A^2} \ = \ m_A^2 \ .
\end{eqnarray}

The one-loop corrected triple Higgs boson coupling, $\lambda_{hhh}$,
 is evaluated as
\begin{eqnarray}
\lambda_{hhh} &=& \frac{\partial^3 V}{\partial h^3} \ [v_u, v_d, 0, 0] \ .
\end{eqnarray}
For convenience, we define ``the deviation of the triple Higgs boson coupling from the SM value",
 $\Delta \lambda_{hhh}/\lambda_{hhh}\vert_{SM}$, as
\begin{eqnarray}
\frac{ \Delta \lambda_{hhh} }{ \lambda_{hhh}\vert_{SM} } &\equiv& 
\frac{ \lambda_{hhh} - \lambda_{hhh}\vert_{SM} }{ \lambda_{hhh}\vert_{SM} } \ .
\end{eqnarray}

\section{Phenomenological Consequences}

\ \ \ We make a numerical analysis on the correlation among 
 the strength of EWPT,
 the decay branching ratio of the Higgs boson into diphoton and
 the triple Higgs boson coupling,
 for a benchmark mass spectrum.
The benchmark is as follows.
For the MSSM sector, 
\begin{eqnarray}
& & \tan \beta \ = \ 15 \ , \ \ \
m_{H^{\pm}} \ = \ 350 \ {\rm GeV} \ , \ \ \ 
\mu \ = \ 200 \ {\rm GeV} \ , \nonumber \\
& & \tilde{M}_{\tilde{t}} \ = \ \tilde{M}_{\tilde{b}} \ = \ 2000 \ {\rm GeV} \label{mssm sample} \ .
\end{eqnarray}
For the $Z_2$-odd sector,
\begin{eqnarray}
& & \mu_{\Phi} \ = \ \mu_{\Omega} \ = \ 550 \ {\rm GeV} \ , \nonumber
\\
& & \bar{m}_{\Phi_{d}} \ = \ \bar{m}_{\Omega^+} \ = \ \bar{m}_{\zeta} \ = \ 1500 \ {\rm GeV} \ , 
\ \ \ \bar{m}_{\eta} \ = \ 2000 \ {\rm GeV} \ , \nonumber 
\\
& & ({\rm A terms, \ B terms}) \ = \ 0 \label{z2odd sample} \ .
\end{eqnarray}
The following two quantities are the free parameters in this analysis:
\begin{eqnarray}
\lambda \ , \ \ \ \ \  m_{0}  \ (&\equiv& \bar{m}_{\Phi_{u}} \ = \ \bar{m}_{\Omega^-}) \label{free} \ .
\end{eqnarray}
We tune the value of the stop mixing term
 to realize $m_h=126$ GeV.

\begin{figure}[htbp]
  \begin{center}
   \includegraphics[width=100mm]{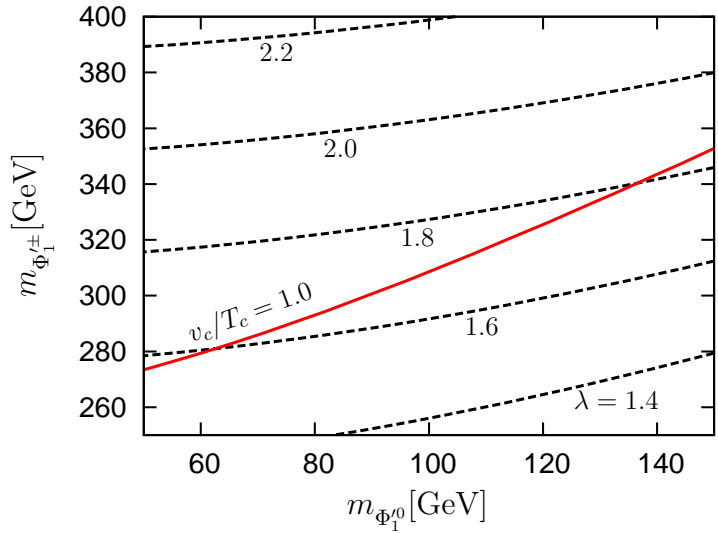}
  \end{center}
  \caption{Contour plot for the coupling constant $\lambda$ (black dashed lines) 
   with a line corresponding to the strength of EWPT $v_C/T_C=1$
   (red solid line),
   on the plane of 
   the mass of the lightest $Z_2$-odd \textit{charged} particle $m_{\Phi^{\prime \, \pm}_1}$
   and the mass of the lightest $Z_2$-odd \textit{neutral} particle $m_{\Phi^{\prime \, 0}_1}$.
  The parameters are fixed according to eqs.~(\ref{mssm sample}) and (\ref{z2odd sample}).
  }
  \label{vctc}
\end{figure}

The results are shown by contour plots on the plane of $m_{\Phi_1^{\prime \, 0}}$ and $m_{\Phi_1^{\prime \, \pm}}$,
 defined respectively as the smallest eigenvalues of the $Z_2$-odd neutral scalar mass matrix in eq.~(\ref{chsc})
 and the $Z_2$-odd neutral scalar mass matrix in eq.~(\ref{nesc}).
Notice that $m_{\Phi_1^{\prime \, 0}}$ and $m_{\Phi_1^{\prime \, \pm}}$ are in one-to-one
 correspondence with $\lambda$ and $m_{0}$ in eq.~(\ref{free}).
In Figure~\ref{vctc}, we show the contour plot for the coupling constant $\lambda$.
The strength of EWPT, $v_C/T_C=1$, is also displayed.
We find that
 strongly first order phase transition, $v_C/T_C \gtrsim 1$, takes places
 with our benchmark mass spectrum
 for $\lambda \gtrsim 1.6$ when $m_{\Phi^{\prime \, 0}_1} \simeq 60$ GeV
 (for $\lambda \gtrsim 1.8$ when $m_{\Phi^{\prime \, 0}_1} \simeq 130$ GeV).
Loop corrections involving light $Z_2$-odd scalars strengthen the order of EWPT.
Hence the lighter the lightest $Z_2$-odd scalar is,
 the smaller value of $\lambda$ we need to realize $v_C/T_C \gtrsim 1$.
We also note that the value of $\lambda$ corresponds to 
 the confinement scale, $\Lambda_H$, of the SUSY SU(2)$_H$ gauge theory in UV.
According to Figure~\ref{figlam}, 
 $\lambda \simeq 1.6$ corresponds to $\Lambda_H \simeq 15$ TeV
 and $\lambda \simeq 1.8$ does to $\Lambda_H \simeq 5$ TeV.

\begin{figure}[htbp]
  \begin{center}
   \includegraphics[width=100mm]{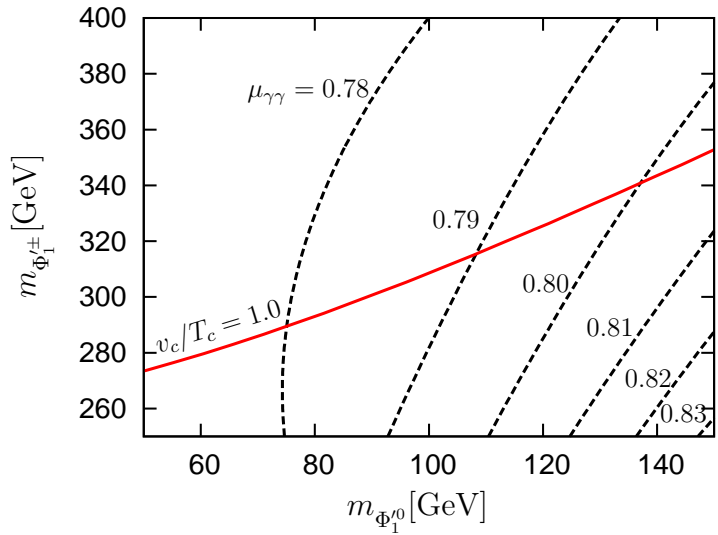}
  \end{center}
  \caption{Contour plot for the ratio of $Br(h \rightarrow \gamma \gamma)$ 
   over the SM value, $\mu_{\gamma \gamma}$ (black dashed lines),
   with a line corresponding to the strength of EWPT $v_C/T_C=1$
   (red solid line), 
   on the plane of
   the mass of the lightest $Z_2$-odd \textit{charged} particle $m_{\Phi^{\prime \, \pm}_1}$
   and the mass of the lightest $Z_2$-odd \textit{neutral} particle $m_{\Phi^{\prime \, 0}_1}$.
  The parameters are fixed according to eqs.~(\ref{mssm sample}) and (\ref{z2odd sample}).
  }
  \label{vctchgg}
\end{figure}

In Figure~\ref{vctchgg}, we combine the contour plot for 
 the ratio of the Higgs-to-diphoton branching ratio over its SM value, 
 $\mu_{\gamma \gamma}$,
 with a line indicating the strength of EWPT, $v_C/T_C=1$.
We find that the Higgs-to-diphoton branching ratio decreases by more than 20 \% 
 with our benchmark mass spectrum when the strongly first order phase transition with $v_C/T_C \gtrsim1$
 is realized.
The deviation of the branching ratio, $\mu_{\gamma \gamma}$,
 exhibits only a mild dependence on $m_{\Phi^{\prime \, 0}_1}$ and $m_{\Phi^{\prime \, \pm}_1}$.
This is because,
 in the sample mass spectrum, 
 the mass of the lightest $Z_2$-odd charged scalar increases with $\lambda$,
 $m_{\Phi^{\prime \, \pm}_1} \sim \lambda v_u$.
Therefore, for loop diagrams contributing to the Higgs-to-diphoton decay,
 the increase in the coupling between the SM-like Higgs boson and the charged scalar
 is cancelled by the increase in the charged scalar mass,
 and thus the deviation of the Higgs-to-diphoton decay is not sensitive to $\lambda$.

\begin{figure}[tbp]
  \begin{center}
   \includegraphics[width=100mm]{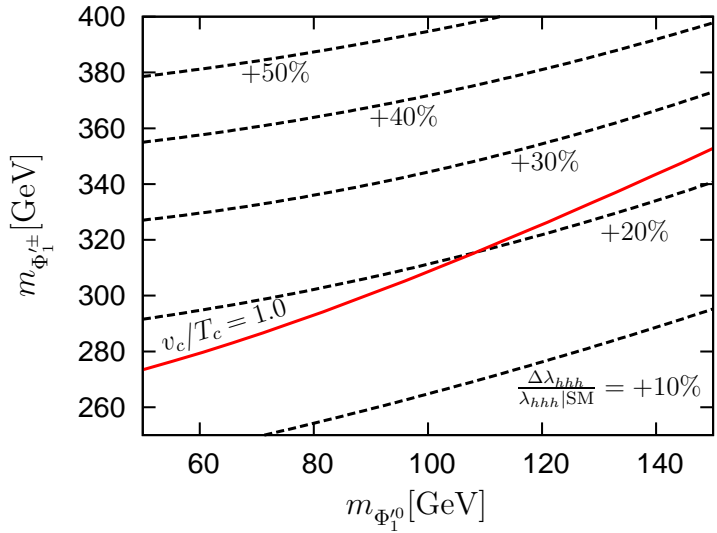}
  \end{center}
  \caption{Contour plot for the deviation of the triple Higgs boson coupling from the SM value, 
   $\Delta \lambda_{hhh}/\lambda_{hhh}\vert_{SM}$
   (black dashed lines),
   with a line corresponding to the strength of EWPT $v_C/T_C=1$
   (red solid line), on the plane of
   the mass of the lightest $Z_2$-odd \textit{charged} particle $m_{\Phi^{\prime \, \pm}_1}$
   and the mass of the lightest $Z_2$-odd \textit{neutral} particle $m_{\Phi^{\prime \, 0}_1}$.
  The parameters are fixed according to eqs.~(\ref{mssm sample}) and (\ref{z2odd sample}).
  }
  \label{vctchhh}
\end{figure}

Finally in Figure~\ref{vctchhh}, we combine the contour plot for 
 the deviation of the triple Higgs boson coupling from the SM value,
 $\Delta \lambda_{hhh}/\lambda_{hhh}\vert_{SM}$,
 with a line indicating the strength of EWPT, $v_C/T_C=1$.
We discover that,
 when the strongly first order EWPT with $v_C/T_C \gtrsim 1$ occurs
 with our benchmark spectrum,
 the triple Higgs boson coupling increases by more than about 20 \% 
 for 150 GeV $> m_{\Phi^{\prime \, 0}_1} >$ 50 GeV.
The strength of EWPT and the deviation of the triple Higgs boson coupling
 are correlated because the same loop corrections involving light $Z_2$-odd scalars
 contribute to both of them.
\\

To summarize, we confirm that sufficiently strong first order EWPT for
 successful EWBG
 can be realized with our benchmark mass spectrum.
In order to have $v_C/T_C \gtrsim 1$,
 we need $\lambda > 1.6$
 provided the lightest $Z_2$-odd neutral scalar is heavier than 50 GeV.
This corresponds to the confinement scale $\Lambda_H$ lower than about 15 TeV.
In the parameter regions where the strongly first order EWPT occurs,
 the Higgs-to-diphoton branching ratio, $Br(h \rightarrow \gamma \gamma)$,
 and the triple Higgs boson coupling, $\lambda_{hhh}$, significantly deviate
 from the SM values.
These are principally due to loop corrections involving light $Z_2$-odd scalars,
 which are also responsible for the strongly first order electroweak phase transition.
With the benchmark mass spectrum,
 $Br(h \rightarrow \gamma \gamma)$ decreases by about 20\%
 and $\lambda_{hhh}$ increases by more than about 20\%,
 both of which may be observed at the future International Linear Collider \cite{peskin, fujii}
 and its $\gamma \gamma$ option \cite{kawada} and the Compact Linear Collider \cite{clic}.
\\

\section{Conclusions}

\ \ \ We have discussed the correlation among the strength of EWPT,
 the Higgs-to-diphoton branching ratio and the triple Higgs boson coupling
 in the extended Higgs sector with large coupling constants and the 126 GeV Higgs boson,
 which emerges as a low-energy effective theory of
 the SUSY SU(2)$_H$ gauge theory with confinement.
In our benchmark mass spectrum,
 the condition of quick sphaleron decoupling for EWBG, $v_C/T_C \gtrsim 1$,
 determines the scale of the Landau pole 
 to be below about 15 TeV, 
 which corresponds to the confinement scale of the SU(2)$_H$ gauge theory.
We have found that the Higgs-to-diphoton branching ratio deviates negatively from the SM prediction
 by about 20\%
 and the triple Higgs boson coupling deviates positively by more than about 20\%.
Such deviations can be observed at future collider experiments.
\\

\acknowledgments

\ \ \ This work was supported in part by Grant-in-Aid for Scientific Research,
Nos. 22244031 (S.K.), 23104006 (S.K.), 23104011 (T.S.) and 24340046 (S.K. and T.S.).
The work of T.Y. was supported in part by a grant of the Japan Society for the Promotion of Science,
 No. 23-3599.


\appendix
\section{One-loop finite temperature effective potential}
We study the electroweak phase transition in the subspace spanned  by $h_d$ and $h_u$,
assuming the other fields do not develop the VEVs.
The nonzero temperature effective potential is 
\begin{align}
V_1(h_d,h_u;T) = \sum_ic_i\frac{T^4}{2\pi^2}I_{B,F}\left(\frac{m_i^2}{T^2}\right),
 \end{align}
where $c_i$ denote the degrees of freedom of the particle species $i$,
$B(F)$ refer to boson (fermion) and $I_{B,F}$ take the form
\begin{eqnarray}
I_{B,F}(a^2) = \int_0^\infty dx~x^2\ln\Big(1\mp e^{-\sqrt{x^2+a^2}}\Big).\label{IBF}
\end{eqnarray}
For a reduction of a computational time, 
we use the fitting functions of $I_{B,F}(a^2)$ that are employed in Ref.~\cite{Funakubo:2009eg}. 
More explicitly,
\begin{eqnarray}
\tilde{I}_{B,F}(a^2)=e^{-a}\sum^N_{n=0}c^{b,f}_na^n,
\label{V1_fit}
\end{eqnarray}
are used, where $c^{b,f}_n$ are determined by the least square method.  For $N=40$, $|I_{B,F}(a^2)-\tilde{I}_{B,F}(a^2)|<10^{-6}$ for any $a$, which is sufficient in our investigation. 

As is well known, the validity of the perturbative expansion would get worse at high temperatures.
The standard prescription for this problem is resummation of the dominant temperature corrections.
Here, we adopt the Parwani's method~\cite{Parwani:1991gq}
in which the both zero and nonzero modes of Matsubara frequencies are resummed.
In this resummation scheme, the particle masses appearing in the one-loop effective potential 
are replaced with the thermally corrected masses.
 
\subsection{Thermal masses}
Let us define the temperature-dependent part of the self energy of a particle $X$ by $\Sigma_X^{(Y)}(T)$,
where $Y$ denotes a particle in the loop.
At the high temperature, $\Sigma_X^{(Y)}(T)$ can be expanded in powers of $m_Y/T(\equiv a_Y)$
\begin{align}
\Sigma_X^{(Y)}(T)
&= C_{XY} \cdot I_B'(a_Y^2) \simeq C_{XY}
	\left[
	\frac{\pi^2}{12}-\frac{\pi}{4}(a_Y^2)^{1/2}
	-\frac{a_Y^2}{16}\left(\ln\frac{a_Y^2}{\alpha_B}-1\right)+\mathcal{O}(a_Y^4)
	\right],
\end{align}
where $C_{XY}$ denotes the coupling constant of $X$ with $Y$ 
and counts the degrees of freedom, $I_B'(a_Y^2)$ is the first derivative of $I_B(a_Y^2)$ 
with respect to $a_Y^2$
and $\ln \alpha_B=2\ln 4\pi-2\gamma\simeq 3.9076$. 
The self energies of $\Omega^\pm$, $\Phi_{u,d}$, $\zeta$ and $\eta$
to leading order in the high temperature expansion are respectively given by
\begin{align}
\Sigma_{\Omega^+}(T) 
&= \Sigma_{\Omega^+}^{(H_d)}(T)
	+\Sigma_{\Omega^+}^{(\Phi_d)}(T)
	+\Sigma_{\Omega^+}^{(\Omega^+)}(T)
	+\Sigma_{\Omega^+}^{(\Omega^-)}(T)
	\nonumber\\
&= \left[\frac{|\lambda|^2}{6}+\frac{|\lambda|^2}{6}+\frac{g_Y^2}{6}-\frac{g_Y^2}{12}\right]T^2, 
\label{Sigma_Omegp}\\
\Sigma_{\Omega^-}(T)
&= \Sigma_{\Omega^-}^{(H_u)}(T)
	+\Sigma_{\Omega^-}^{(\Phi_u)}(T)
	+\Sigma_{\Omega^-}^{(\Omega^-)}(T)
	+\Sigma_{\Omega^-}^{(\Omega^+)}(T) \nonumber\\ 
&= \left[\frac{|\lambda|^2}{6}+\frac{|\lambda|^2}{6}+\frac{g_Y^2}{6}-\frac{g_Y^2}{12}\right]T^2, 
\label{Sigma_Omegm}\\
\Sigma_{\Phi_d}(T) 
&= \Sigma_{\Phi_d}^{(H_d)}(T)
	+\Sigma_{\Phi_d}^{(\Phi_d)}(T)
	+\Sigma_{\Phi_d}^{(\Phi_u)}(T) 
	+\Sigma_{\Phi_d}^{(\eta)}(T) 	
	\nonumber\\
&= \left[\frac{|\lambda|^2}{6}+\frac{g^2+g_Y^2}{16}-\frac{g_Y^2}{24}
	+\frac{|\lambda|^2}{6}\right]T^2, \label{Sigma_Phid}\\
\Sigma_{\Phi_u}(T) 
&= \Sigma_{\Phi_u}^{(H_u)}(T)
	+\Sigma_{\Phi_u}^{(\Phi_u)}(T)
	+\Sigma_{\Phi_u}^{(\Phi_d)}(T) 
	+\Sigma_{\Phi_u}^{(\zeta)}(T) 	
	\nonumber\\
&= \left[\frac{|\lambda|^2}{6}+\frac{g^2+g_Y^2}{16}
	-\frac{g_Y^2}{24}+\frac{|\lambda|^2}{6}\right]T^2, \label{Sigma_Phid}\\
\Sigma_{\zeta}(T) 
&=\Sigma_{\zeta}^{(H_d)}(T)+\Sigma_{\zeta}^{(\Phi_u)}(T)  \nonumber\\
&=\left[\frac{|\lambda|^2}{6}+\frac{|\lambda|^2}{6}\right]T^2, \label{Sigma_zeta}\\
\Sigma_{\eta}(T) 
&=\Sigma_{\eta}^{(H_u)}(T)+\Sigma_{\eta}^{(\Phi_d)}(T)  \nonumber\\
&=\left[\frac{|\lambda|^2}{6}+\frac{|\lambda|^2}{6}\right]T^2.\label{Sigma_eta}
\end{align}
Here, we only show the Higgs boson loop contributions.
The contributions of their superpartners are half of them.

Note that $I_B'(a^2)$ is Boltzmann suppressed for $a=m/T>1$ as
\begin{align}
I_B'(a^2) \simeq \frac{1}{2}\sqrt{\frac{\pi a}{2}}e^{-a}\left[1+\frac{3}{8a}+\cdots\right].
\end{align}
Therefore, we remove the $T^2$ corrections of $\Sigma_X^{(Y)}(T)$ from
(\ref{Sigma_Omegp})-(\ref{Sigma_eta}) in such a large mass region.
In our analysis, in addition to the gauge bosons, 
$\Omega^-$, $\Phi_u$ and $H_u$  
are potentially light enough to contribute to the screening effects
since $\bar{m}_{\Omega^-}=\bar{m}_{\Phi_u}=$ 50 GeV.
It turns out that $|\bar{m}_{H_u}|>200$ GeV in most parameter space. 
So the thermal resummation in our EWPT study are done by the following replacements
\footnote{Even though the screening effect of $H_u$ is taken into account, 
the critical line of $v_c/T_c=1$ is not significantly changted. 
However, the maximal value of $v_c/T_c$ cannot exceed 1.5 in this case.
}
\begin{align}
\bar{m}_{\Omega^-}^2 &\to \bar{m}_{\Omega^-}^2 
	+\Sigma_{\Omega^-}^{(\Phi_u)}(T)+\Sigma_{\Omega^-}^{(\Omega^-)}(T)
	+\Sigma_{\Omega^-}^{(\rm gauge)}(T), \\
\bar{m}_{\Phi_u}^2 & \to \bar{m}_{\Phi_u}^2 + \Sigma_{\Phi_u}^{(\Phi_u)}(T)
	+\Sigma_{\Phi_u}^{(\rm gauge)}(T),
\end{align}
where $\Sigma_{\Omega^-, \Phi_u}^{(\rm gauge)}(T)$ denote the gauge boson contributions.
\\

\end{document}